\documentclass[aps,pra,twocolumn,superscriptaddress,10pt,tightenlines,nofootinbib]{revtex4}
\usepackage{amsmath,amsthm,graphicx,amssymb,verbatim,listings,mathrsfs}
\usepackage{verbatimbox}
\usepackage[usenames, dvipsnames]{color}
\usepackage[ pdftex, plainpages = false, pdfpagelabels,
                 pdfpagelayout = useoutlines,
                 bookmarks,
                 bookmarksopen = true,
                 bookmarksnumbered = true,
                 breaklinks = true,
                 linktocpage=all,
                 pagebackref=false,
                 colorlinks = true,
                 linkcolor = BrickRed,
                 urlcolor  = blue,
                 citecolor = BrickRed,
                 anchorcolor = green,
                 hyperindex = true,
                 hyperfigures
                 ]{hyperref}
\usepackage{xifthen} 

\newcommand{\tr}{\hbox{tr}}

\newcommand{\ket}[1]{{\ensuremath{\left| #1 \right\rangle}}}
\newcommand{\bra}[1]{{\ensuremath{\left\langle #1 \right|}}}
\newcommand{\braket}[2]{{\ensuremath{\left\langle #1 \middle| #2
      \right\rangle}}}
\newcommand{\ketbra}[2]{{\ensuremath{\left| #1 \middle\rangle \middle\langle #2
      \right|}}}

\newcommand{\sympdes}{{\ensuremath{\mathscr{S}}}}
\newcommand{\stabgp}{{\ensuremath{PSU(3,3)}}}

\newcommand{\arxiv}[2][]{\ifthenelse{\isempty{#1}}{\href{http://arxiv.org/abs/#2}{{\tt arXiv:\allowbreak{}#2}}} {\href{http://arxiv.org/abs/#2}{{\tt arXiv:\allowbreak{}#2 [#1]}}}}

\newcommand{\booktitle}{\textsl}
\newcommand{\hrefdoi}[2]{\href{https://dx.doi.org/#1}{#2}}

\begin{document}

\title{Geometric and Information-Theoretic Properties of the Hoggar Lines}

\author{Blake C.\ Stacey}
\affiliation{Department of Physics,
  University of Massachusetts Boston, 100 Morrissey Blvd.,
  Boston, MA 02125, United States}

\date{\today}

\begin{abstract}
We take a tour of a set of equiangular lines in eight-dimensional
Hilbert space.  This structure defines an informationally complete
measurement, that is, a way to represent all quantum states of
three-qubit systems as probability distributions.  Investigating the
shape of this representation of state space yields a pattern of
connections among a remarkable spread of mathematical constructions.
In particular, studying the Shannon entropy of probabilistic
representations of quantum states leads to an intriguing link between
the questions of real and of complex equiangular lines.  Furthermore,
we will find relations between quantum information theory and
mathematical topics like octonionic integers and the 28 bitangents to
a quartic curve.
\end{abstract}

\maketitle

\section{Introduction}

A set of \emph{equiangular lines} is a collection of lines such that
the angle made by each pair of lines is equal.  These arrangements can
be defined in real vector space $\mathbb{R}^d$ or in complex vector
space $\mathbb{C}^d$.  The outstanding question is what the
\emph{maximum} size of such a set can be, as a function of the
dimension $d$.  This question is relevant to quantum physics, because
the complex case corresponds to a particular type of \emph{quantum
  measurement} with important properties~\cite{Zauner99, Renes04,
  Quantumness, Scott10, LieAlgebra1, Voldemort, RMP, LieAlgebra2,
  Fuchs2014, ConicalDesigns, RCF-SIC, Zhu2016, stacey-sporadic}.  One
moral of our story will be that the real and the complex versions of
the equiangular line question can intertwine in unexpected ways.

A symmetric, informationally complete, positive-operator valued
measure---a SIC-POVM, or just a SIC---is a set of $d^2$ vectors
$\ket{\psi_j}$ in a $d$-dimensional complex Hilbert space whose inner
products satisfy
\begin{equation}
\left|\braket{\psi_j}{\psi_k}\right|^2
 = \frac{d\delta_{jk} + 1}{d+1}.
\label{eq:SIC-definition}
\end{equation}
It is often convenient to work with the rank-1 projection operators
defined from these states,
\begin{equation}
\Pi_j = \ketbra{\psi_j}{\psi_j}.
\end{equation}
When rescaled by the dimension, these operators sum to the identity:
\begin{equation}
\sum_j \frac{1}{d} \Pi_j = I.
\end{equation} 
Therefore, with this scaling, the operators $\Pi_j$ can serve as the
\emph{effects} that comprise a general quantum measurement, or POVM.
The index $j$ labels the possible outcomes of an experiment that can,
in principle, be carried out.  It follows from
Eq.~(\ref{eq:SIC-definition}) that such a measurement is
\emph{informationally complete} (IC).  Given a probability
distribution over the outcomes of an IC measurement, we can compute
the probabilities for the outcomes of any other measurement.  The
\emph{symmetric} IC POVMs make the calculations that interrelate
different experiments take on a remarkably simple form~\cite{RMP,
  Fuchs2014}.

One can prove that no more than $d^2$ states in a $d$-dimensional
Hilbert space can be equiangular.  That is, the largest set of states
for which
\begin{equation}
\left|\braket{\psi_j}{\psi_k}\right|^2
 = \alpha
\label{eq:C-equiangular}
\end{equation}
whenever $j \neq k$ has size $d^2$.  In turn, for a maximal set the
value of~$\alpha$ is fixed by the dimension; it must be $1/(d+1)$.
So, a SIC is a maximal equiangular set in~$\mathbb{C}^d$; the question
is whether they can be constructed for \emph{all} values of the
dimension.  Despite a substantial number of exact solutions, as well
as a longer list of high-precision numerical solutions~\cite{Scott10,
  ConicalDesigns, RCF-SIC}, the problem remains open.

The real vector space analogue to Eq.~(\ref{eq:C-equiangular}) can be
expressed in terms of the Euclidean inner product $\langle \cdot,
\cdot \rangle$ in~$\mathbb{R}^n$.  An equiangular set of unit vectors
$\hat{v}_j$ satisfies
\begin{equation}
\langle \hat{v}_j, \hat{v}_k \rangle = \pm \alpha
\end{equation}
for all $j \neq k$.  Again, one can find an upper bound for the
possible size of such a set.  In Euclidean space $\mathbb{R}^d$, an
equiangular set can contain at most
\begin{equation}
N = \binom{d+1}{2}
\end{equation}
lines.  However, unlike the bound of $d^2$ in the complex case, it is
known that this bound is \emph{not} reached in all
dimensions~\cite{sloane2015, greaves2016}.  For example,
in~$\mathbb{R}^7$, one can construct $\binom{8}{2} = 28$ equiangular
lines, but this is also the best that one can do in~$\mathbb{R}^8$ and
in~$\mathbb{R}^9$.
 
The general plan for this essay is as follows.  In Sections
\ref{sec:sicrep} and \ref{sec:constructing}, we will see how SICs
furnish a probabilistic representation of quantum state space, and we
will introduce the particular SICs that will be our main focus of
interest.  Recent work by Szymusiak and S\l{}omczy\'nski has
demonstrated the importance of these SIC solutions for understanding
the \emph{informational power} of quantum
measurements~\cite{Szymusiak2015}.  We will touch on these
calculations, in the context of extremizing the Shannon entropy of
probabilistic representations for quantum states.  These results
connect complex geometry, finite group theory and information theory.
Next, in Section~\ref{sec:qbic-eq}, we will use the special properties
of those SICs to simplify the equations that indicate the shape of
quantum state space.  Section~\ref{sec:triple} will build on that
development, showing how the way SIC vectors embed into Hilbert space
leads to combinatorial design theory.  Teasing out the structures that
arise from an eight-dimensional SIC leads to an unforeseen connection
between the real and the complex versions of the equiangular lines
problem.

Section~\ref{sec:twin} will study the pairing
of two separate eight-dimensional SICs, a pattern of interlocking
geometrical relationships that will lead, in
Section~\ref{sec:twin-design}, to another application of combinatorial
design theory.  The results will translate to probability and
information theory in Section~\ref{sec:PP}, where we will see what
they imply for the problem of distinguishing the consequences of
different quantum-mechanical hypotheses.  Investigating this further,
we will arrive in Section~\ref{sec:deeper} at another connection
between real and complex equiangular lines.

\section{SIC Representations of Quantum States}
\label{sec:sicrep}
In the textbook way of doing quantum theory, a quantum state for a
system is a positive semidefinite operator $\rho$ with unit trace.
For a $d$-dimensional system (a \emph{qudit}), $\rho$ can be written
as a $d\times d$ matrix of complex numbers.  The set of all valid
\emph{density matrices} $\rho$ is a convex set whose extreme points
are the rank-1 projection operators.  These extreme points are also
known as \emph{pure states}; states that are not pure are designated
\emph{mixed.}

Given a density matrix $\rho$ and a POVM $\{E_i\}$, we find the
probability of outcome $i$ by using the \emph{Born rule}:
\begin{equation}
\hbox{Prob}(i) = \tr(\rho E_i).
\end{equation}
If the POVM is informationally complete, we can reconstruct $\rho$
entirely from these probabilities.  In the case of a SIC~\cite{RMP},
we can say that
\begin{equation}
\rho = \sum_i \left((d+1)p(i) - \frac{1}{d}\right) \Pi_i,
\end{equation}
where
\begin{equation}
p(i) = \frac{1}{d} \tr(\rho \Pi_i).
\end{equation}
We will call the probability distribution $p(i)$ the \emph{SIC
  representation} of the quantum state $\rho$.

Let us suppose we have a SIC solution for some dimension $d$.  (In the
next section, we will examine some examples in detail.)  A state
is pure if and only if its SIC representation satisfies the following
two conditions.  First, it must meet the quadratic constraint
\begin{equation}
\sum_j p(j)^2 = \frac{2}{d(d+1)}.
\label{eq:quadratic-constraint}
\end{equation}
Second, it must satisfy the \emph{QBic equation,}
\begin{equation}
\sum_{jkl} C_{jkl} p(j)p(k)p(l) = \frac{d + 7}{(d+1)^3},
\label{eq:qbic-general}
\end{equation}
where we have introduced the \emph{triple products,}
\begin{equation}
C_{jkl} = \hbox{Re}\,\tr(\Pi_j\Pi_k\Pi_l).
\end{equation}
If two or more indices are equal, this reduces to
\begin{equation}
\tr(\Pi_j\Pi_k) = \frac{d\delta_{jk}+1}{d+1}.
\end{equation}
The set of all valid states is the convex hull of the probability
distributions that satisfy Eqs.~(\ref{eq:quadratic-constraint}) and
(\ref{eq:qbic-general}).

The quadratic constraint (\ref{eq:quadratic-constraint}) has a
considerably simpler structure than the QBic equation, so we
investigate the former first.  One important consequence is an upper
bound on the number of entries in~$p(i)$ that can equal
zero~\cite{appleby2011}.  Normalization implies that
\begin{equation}
1 = \left(\sum_i p(i)\right)^2.
\end{equation}
Writing $n_0$ for the number of zero-valued entries in~$p(i)$, and
applying the Cauchy--Schwarz inequality,
\begin{equation}
\left(\sum_i p(i)\right)^2
 \leq (d^2 - n_0) \sum_{\{i | p(i) \neq 0\}} p(i)^2.
\end{equation}
Consequently,
\begin{equation}
1 \leq (d^2 - n_0) \frac{2}{d(d+1)}.
\end{equation}
Rearranging this, we find that
\begin{equation}
n_0 \leq \frac{d(d-1)}{2} = \binom{d}{2}.
\end{equation}
When this bound was first derived, it was conjectured that one could
improve upon it~\cite{appleby2011}.  This bound can be reached in
dimension 3.  Note that when $d = 3$, the binomial coefficient
$\binom{d}{2}$ reduces to~$d$.  It was conjectured that the true bound
would turn out to be $d$ in general.  However, this is not the
case~\cite{stacey-qutrit}.  In this paper, we will find examples in~$d
= 8$ where the number of zeros is $\binom{8}{2} = 28$.  Therefore, the
bound deduced from the Cauchy--Schwarz inequality is the best one
possible in general.

Since we have probability distributions, we can compute Shannon
entropies.  Of particular interest are the pure states which extremize
the Shannon entropy of their SIC representations.  It turns out (and
the proof is not too long) that the pure states which \emph{maximize}
the Shannon entropy of their SIC representations are the SIC
projectors $\{\Pi_i\}$ themselves.\footnote{I first learned of this
  from unpublished notes by Huangjun Zhu, written in 2013.}

What about \emph{minimizing} the Shannon entropy?  Imagine a
probability distribution, not necessarily one corresponding to a
quantum state.  Under the constraint that $\sum_i
p(i)^2$ is fixed,
\begin{equation}
\sum_i p(i)^2 = \frac{1}{N},
\end{equation}
then it can be shown~\cite{Szymusiak2015} that the distributions of
minimum entropy take the form
\begin{equation}
\left(\frac{1}{N},\cdots,\frac{1}{N},0,\cdots,0\right).
\label{eq:Shannon-minimizer}
\end{equation}
Exactly $N$ entries are nonzero, and the others all vanish.  If we
take
\begin{equation}
N = \frac{d(d+1)}{2} = \binom{d+1}{2},
\end{equation}
then we see that a probability distribution with $N$ nonvanishing,
uniform entries is a pure state that minimizes the Shannon
entropy---\emph{provided} that it corresponds to a valid pure state.
In other words, the minimizers we seek are those permutations of
Eq.~(\ref{eq:Shannon-minimizer}) that satisfy the QBic equation.

\section{Constructing SICs Using Groups}
\label{sec:constructing}
All known SICs have an additional kind of symmetry, above and beyond
their definition: They are \emph{group covariant.}  Each SIC can be
constructed by starting with a single vector, known as a
\emph{fiducial} vector, and acting upon it with the elements of some
group.  It is not known whether or not a SIC must be group covariant.
Possibly, because group covariance simplifies the search
procedure~\cite{Scott10, RCF-SIC}, the fact that we only know of
group-covariant SICs is merely an artifact.  (However, we do have a
proof that all SICs in~$d = 3$ are group
covariant~\cite{Hughston2016}.)

In all cases but one, namely the Hoggar SIC we will define below, the
group that generates a SIC from a fiducial is an instance of a
\emph{Weyl--Heisenberg group.}  We can define this group as
follows. First, fix a value of~$d$, and let $\omega_d = e^{2\pi i /
  d}$. Then, construct the shift and phase operators
\begin{equation}
X\ket{j} = \ket{j+1},\ Z\ket{j} = \omega_d^j \ket{j},
\end{equation}
where the shift is modulo $d$.  The elements of the Weyl--Heisenberg
group in dimension $d$ are products of powers of $X$ and $Z$, together
with phase factors that depend on the dimension.  For many purposes,
those phase factors can be neglected.

In $d = 2$---that is, for a system comprising a single qubit---a SIC
is simply a tetrahedron, inscribed in the Bloch sphere~\cite{Renes04}.
(This configuration was described by Feynman, in a 1987 festschrift
for Bohm~\cite{Feynman1987}.)  Let $r$ and $s$ be signs, and let
$\sigma_x$, $\sigma_y$ and $\sigma_z$ denote the Pauli matrices.
Then, the four pure states
\begin{equation}
\Pi_{r,s} = \frac{1}{2} \left(I + \frac{1}{\sqrt{3}}
                             (r\sigma_x + s\sigma_y + rs\sigma_z)
                       \right)
\label{eq:tetrahedron1}
\end{equation}
define a tetrahedron.  Each point $(x,y,z)$ lying within the unit ball
(\emph{Bloch ball}) defines a valid quantum state.  The SIC
representation of this state is the probability vector whose components are
\begin{equation}
p(r,s) = \frac{1}{4}
 + \frac{\sqrt{3}}{12}
   \left(sx + ry + srz\right).
\end{equation}
Given the tetrahedron (\ref{eq:tetrahedron1}), we can define another,
related to the first by inversion.  Together, the two tetrahedra form
a stellated octahedron, inscribed in the Bloch sphere.  The SIC
representations of the vertices of the second tetrahedron are the
vector
\begin{equation}
\left(0, \frac{1}{3}, \frac{1}{3}, \frac{1}{3}\right)
\end{equation}
and its permutations.

In what follows, we will make substantial use of two SICs.  One of
them is the \emph{Hesse SIC} in $d = 3$, which we construct by
applying the Weyl--Heisenberg group to the fiducial
\begin{equation}
\ket{\psi_0^{\rm (Hesse)}} 
 = \frac{1}{\sqrt{2}} (0, 1, -1)^{\rm T}.
\label{eq:hesse-fiducial}
\end{equation}
The other lives in $d = 8$ and is designated the \emph{Hoggar SIC.}
(The construction was first devised by Hoggar~\cite{Hoggar1981,
  Hoggar1998} by starting with 64 nonequiangular diagonals through the
vertices of a quaternionic polytope, which become 64 equiangular lines
when converted to complex space.  Hoggar's result was among the first
discoveries of a maximal set of complex equiangular
lines~\cite[pp.~731--33]{samizdat}.)  Actually, we have multiple
choices of fiducial in this case, yielding distinct sets of $d^2 = 64$
states.  However, all of these sets have the same symmetry group, and
they are equivalent to one another up to unitary or antiunitary
transformations.  For brevity, then, we will refer to ``the'' Hoggar
SIC~\cite{zhu-thesis}.

A fiducial for the Hoggar SIC~\cite{jedwab2015} can be written as follows:
\begin{equation}
\ket{\psi_0} \propto (-1+2i, 1, 1, 1,
                      1, 1, 1, 1)^{\rm T}.
\label{eq:hoggar-fiducial}
\end{equation}
Upon this, we act with the elements of the group that is the tensor
product of three copies of the $d = 2$ Weyl--Heisenberg group:
\begin{equation}
k = (k_0,k_1,\ldots,k_5),\ 
D_k = X^{k_0} Z^{k_1}
 \otimes X^{k_2} Z^{k_3}
 \otimes X^{k_4} Z^{k_5}.
\label{eq:displacement-operators}
\end{equation}

Given a tetrahedral SIC, we can define a SIC representation of state
space.  Minimizing the Shannon entropy over pure states, as we
discussed earlier, yields the four states of the counterpart
tetrahedron.  Performing the same procedure with the Hesse SIC, we
find that the pure states that minimize the Shannon entropy are twelve
in number.  They form a complete set of \emph{Mutually Unbiased
  Bases}~\cite{stacey-qutrit}.  It is natural to ask what happens
similarly for the Hoggar SIC; we will investigate this in a later
section.

\section{Simplfying the QBic Equation}
\label{sec:qbic-eq}
A quantum system for which $d = 3$ is known as a \emph{qutrit.}  In $d
= 3$, we can simplify the QBic equation~(\ref{eq:qbic-general})
considerably, using the Hesse SIC~\cite{Tabia12, TabiaAppleby13,
  stacey-qutrit}.  In the Hesse SIC representation of qutrit state
space, the QBic equation can be reduced to
\begin{equation}
\sum_i p(i)^3 - 3 \sum_{(ijk) \in S} p(i) p(j) p(k) = 0,
\label{eq:qbic-tabia}
\end{equation}
where list $S$ is a set of index triples $(ijk)$ which can be
constructed as the lines in a discrete affine plane of nine
points~\cite{Tabia12, TabiaAppleby13, stacey-qutrit}.  This fact is
quite handy when working with qutrit states, and it is a consequence
of the triple products of the Hesse SIC states taking a simple form.
In turn, the structure of the triple products simplifies because the
Hesse SIC has the property that its symmetry group acts \emph{doubly
  transitively.}  This is a kind of symmetry beyond the definition of
a SIC and beyond group covariance:  Using unitary operators that map
the Hesse SIC to itself, we can send any pair of states in the Hesse
SIC to any other.

Zhu has proved~\cite{zhu2015} that the only SICs whose symmetry groups
act doubly transitively are the tetrahedral SICs in~$d = 2$, the Hesse
SIC in~$d = 3$ and the Hoggar SIC in~$d = 8$.  In $d = 2$, the QBic
equation simplifies so far that it becomes redundant, and the
quadratic constraint is sufficient to define the state space.  As we
have seen, the QBic equation also simplifies for the Hesse SIC, in a
way that brings discrete geometry into the picture.  It is reasonable
to guess that something similar will happen in dimension $d = 8$.

When in dimension $d = 8$, using the Hoggar SIC, the number
of distinct values the $C_{jkl}$ take in this case is quite small:
When the three indices are different, they can only be $0$ or $\pm 1/27$.

Let $S_+$ denote the set of index triples $(jkl)$ for which $C_{jkl} =
1/27$, and likewise, let $S_-$ denote that set for which $C_{jkl} =
-1/27$.  We cull duplicates from these lists, so that, for example, if
$(jkl)$ belongs in~$S_+$, we do not also include its permutations
$(kjl)$, $(lkj)$ and so on.  The sizes of these sets are
\begin{equation}
|S_+| = 16128 = 2^8 3^2 7,\ |S_-| = 4032 = 2^6 3^2 7.
\end{equation}

Simplifying the QBic equation (\ref{eq:qbic-general}) for the special
case of the Hoggar SIC proceeds by fairly straightforward algebra.
The only bit of moderate cleverness required is a rearrangement by
means of normalization:
\begin{align}
\sum_j p(j)^2 \sum_{l\neq j} p(l)
 &= \sum_j p(j)^2 \left[1 - p(j)\right] \nonumber\\
 &= \sum_j p(j)^2 - \sum_j p(j)^3.
\end{align}
The result of these manipulations is that a pure state must satisfy
\begin{equation}
\sum_j p(j)^3
 + \frac{1}{3} \left[\sum_{S_+} p(j)p(k)p(l)
   - \sum_{S_-} p(j)p(k)p(l)
   \right]
 =
 \frac{11}{648}.
\label{eq:qbic-hoggar-boxed}
\end{equation}
The remaining challenge is to characterize the sets $S_+$ and $S_-$.

\section{Triple Products and Combinatorial Designs}
\label{sec:triple}
Group covariance tells us that any $C_{jkl}$ can be written as
$C_{0mn}$ for some $m$ and $n$.  This implies a $d^2$-fold degeneracy
among the triple products.  In our case, we know that the sizes
of~$S_+$ and $S_-$ must be multiples of~64.  And, in fact,
\begin{equation}
|S_+| = 64 \cdot 4 \cdot 3^2 7,\ |S_-| = 64 \cdot 3^2 7.
\end{equation}
In forming the triple product $C_{0mn}$, we have
\begin{equation}
\binom{63}{2} = \frac{63 \cdot 62}{2} = 1953
\end{equation}
ways of choosing the subscripts $m$ and $n$.  We find that
\begin{equation}
\frac{|S_-|}{64} = \frac{1}{31} \binom{63}{2},
\ \frac{|S_+|}{64} = \frac{4}{31} \binom{63}{2}.
\end{equation}

It is now time to go into the group theory of SIC structures in more
detail.  We define the \emph{multipartite Weyl--Heisenberg
  group} in a prime-power dimension $p^n$ to be the tensor product
of~$n$ copies of the Weyl--Heisenberg group in dimension $p$.  The
\emph{Clifford group} in dimension $p^n$ is the group of unitaries
that stabilize the multipartite Weyl--Heisenberg group.  The order of
the Clifford group~\cite{zhu2015} is
\begin{equation}
p^{n^2 + 2n} \prod_{j=1}^n (p^{2j} - 1).
\end{equation}
Therefore, in dimension $8 = 2^3$, the Clifford group has order
\begin{align}
2^{3^2 + 2\cdot3} \prod_{j=1}^3 (2^{2j} - 1)
 &= 2^{15} \cdot 3 \cdot 15 \cdot 63
 = 2^{15} \cdot 3^4 \cdot 5 \cdot 7 \nonumber\\
 &= 2^9 \cdot 3^2 \cdot (|S_+| + |S_-|).
\end{align} 

The symmetry group of the Hoggar SIC is a subgroup of the Clifford
group with order
\begin{equation}
64 \cdot 6048 = 2^{11} \cdot 3^3 \cdot 7
 = 24|S_+| = 96|S_-|.
\label{eq:hoggar-sym-gp-order}
\end{equation}
The factor of $64 = 2^6$ comes from the triple-qubit Pauli group.
Take any vector from the Hoggar SIC, and consider those unitaries in
the symmetry group that leave that vector fixed while permuting the
others.  These form the stabilizer subgroup of that vector.  For any
vector in the Hoggar SIC, the stabilizer subgroup is isomorphic to the
projective special unitary group $\stabgp$, which has 6048 elements.
This explains the other factor in
Eq.~(\ref{eq:hoggar-sym-gp-order}).\footnote{If one constructs the
  Hoggar SIC as Zhu does, then its fiducial vector's stabilizer group
  is generated by his unitary operators $U_7$ and $U_{12}$.  Construct
  the new unitaries $U_a = U_{12} U_7$ and $U_b = U_{12}^2$.  These
  satisfy the relations for the generators of~$\stabgp$ as presented
  in the \booktitle{Atlas of Finite Group
    Representations}~\cite{AoFGR}.  Also, the conjugacy classes in
  Zhu's Table 10.1 can be matched with those for~$\stabgp$ computed,
  for example, using the \textsc{gap} software~\cite{GAP4}.  The group
  $\stabgp$, as well as the stabilizer groups for the other
  doubly-transitive SICs, can all be constructed from the
  \emph{octavian integers}~\cite{stacey-sporadic, conway-smith}.  As
  Baez notes, ``Often you can classify some sort of gizmo, and you get
  a beautiful systematic list, but also some number of
  exceptions. Nine times out of 10 those exceptions are related to the
  octonions''~\cite{baez-plus}.}

Let $N_k^+$ be the number of triples in the set $S_+$ that contain the
value $k$, and likewise for $N_k^-$ and $S_-$.  One finds that
\begin{equation}
N_k^- = 189,\ N_k^+ = 756\ \forall k.
\end{equation}
These values factorize as
\begin{equation}
N_k^- = 3^3 \cdot 7,\ N_k^+ = 2^2 \cdot 3^3 \cdot 7.
\end{equation}
Furthermore, if we let $N_{kl}^\pm$ denote the number of triples in
$S_+$ (respectively, $S_-$) that contain the pair $(k,l)$, we obtain
\begin{equation}
N_{kl}^- = 6,\ N_{kl}^+ = 24,\ \forall k,l.
\end{equation}

This leads us into \emph{combinatorial design theory.}  A
\emph{balanced incomplete block design} (BIBD) is a collection of $v$
points and $b$ blocks, such that there are $k$ points within each
block, and $r$ blocks contain any given point.  Consistency requires
that
\begin{equation}
bk = vr.
\end{equation}
The final parameter, $\lambda$, specifies the number of blocks
containing any two specific points.  This constant must satisfy
\begin{equation}
\lambda(v-1) = r(k-1).
\end{equation}
In a \emph{symmetric design,} $b = v$, and so $r = k$.  Any two blocks
meet in the same number of points, and that number is $\lambda$.
Ryser's theorem~\cite{ryser1963} establishes that this is an
if-and-only-if relationship.

The set $S_-$ contains 4032 ``blocks,'' where each block is made of
three points drawn from a set of 64 possibilities.  We found earlier
that each point occurs in 189 different blocks, and that each pair of
points occurs in 6 different blocks.  Therefore, $S_-$ is a BIBD with
\begin{equation}
v = 64,\ b_- = |S_-| = 4032,\ k = 3,\ r_- = 189,\ \lambda_- = 6.
\end{equation}
Likewise, $S_+$ is a BIBD with
\begin{equation}
v = 64,\ b_+ = |S_+| = 16128,\ k = 3,\ r_+ = 756,\ \lambda_+ = 24.
\end{equation}
Referring back to Eq.~(\ref{eq:hoggar-sym-gp-order}), we have that
\begin{equation}
b_\pm = |S_\pm| = \frac{6048v}{4\lambda_\mp}
 = \frac{6048v}{4} \frac{\lambda_\pm}{\lambda_-\lambda_+}
 = \frac{6048v}{576} \lambda_\pm = 672\lambda_\pm.
\end{equation}

Zhu proves that the Hoggar SIC is ``doubly transitive,'' \emph{i.e.,}
for any distinct pair of vectors, there is a symmetry operation that
takes it to any other distinct pair~\cite{zhu2015}.  This has
implications for the structure coefficient matrices $C_i$, defined by
\begin{equation}
(C_i)_{jk} = C_{ijk}.
\end{equation}
Group covariance means that
\begin{equation}
C_{ijk} = C_{0j'k'}
\end{equation}
for some $j'$ and $k'$.  So, the entries in all the matrices $\{C_i\}$
are elements of the matrix $C_0$.  The additional requirement that the
action of the symmetry group is doubly transitive means that
if we want to understand the triple products $C_{ijk}$, we only need
to look at $C_{01k}$, because any triple of distinct indices $(ijk)$
can be mapped to some $(01k')$, leaving the triple product invariant.

We expect to see some values occur in sets of six, or multiples of
six.  Why?  Because the triple product function is completely
symmetric:
\begin{equation}
C_{ijk} = C_{jki} = C_{kij} = C_{jik} = C_{kji} = C_{ikj}.
\end{equation}
By applying unitaries in the symmetry group, we can turn the first
pair of indices into $ij$ across the board:
\begin{equation}
C_{ijk} = C_{ij\sigma_1(i)} = C_{ij\sigma_2(j)}
 = C_{ij\sigma_3(k)} = C_{ij\sigma_4(i)} = C_{ij\sigma_5(j)}.
\end{equation}
Here, the $\{\sigma_1,\ldots,\sigma_5\}$ are permutations of the set
of indices $\{0,\ldots,63\}$.  They are defined by relations of the
form
\begin{equation}
\sigma_1(j) = i,\ \sigma_1(k) = j.
\end{equation}
Unless these permutations happen to align in such a way that, for
example, $\sigma_4(i) = \sigma_5(j)$, we will have six elements in the
$j^{\rm th}$ row of the matrix $C_i$, all equal.

\begin{figure}[h]
\includegraphics[width=5cm]{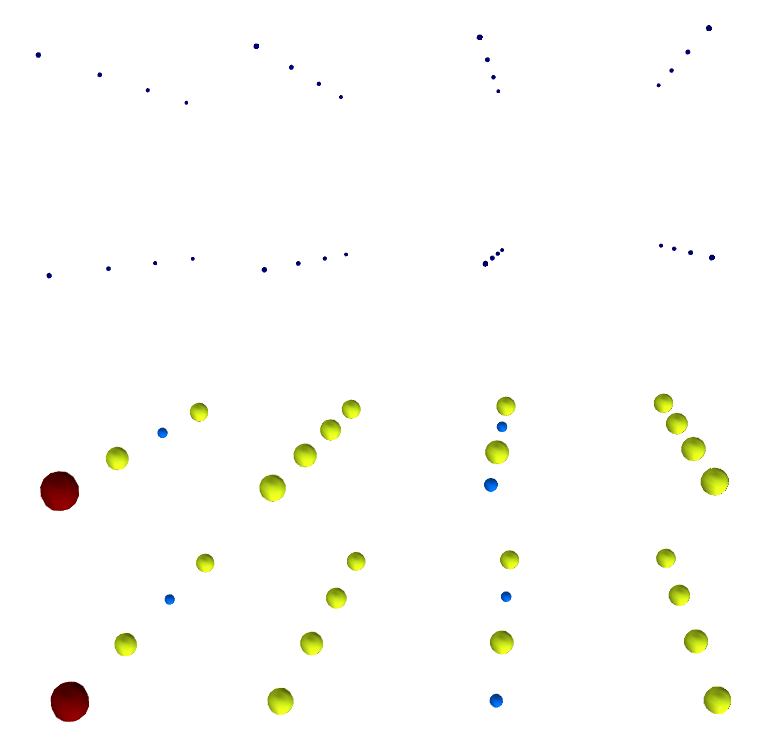}
\caption{\label{fig:triple-product-cube} Visual representation of
  $C_{01k}$ for the Hoggar SIC.  Small dots indicate $C_{01k} = 0$.
  Large spheres (red) indicate the trivial value, $C_{01k} = 1/9$.
  Intermediate spheres (yellow) indicate $C_{01k} = 1/27$, and the six
  slightly smaller spheres (blue) stand for $C_{01k} = -1/27$.}
\end{figure}

Explicit computation bears this idea out.  We need the values of
$C_{01k}$, where the subscripts ``0'' and ``1'' refer to the first and
second projectors in the ordering defined by
Eq.~(\ref{eq:displacement-operators}).  Note that two
entries will be the trivial value, $1/(d+1)$.
We can display the results by arranging them in a $4 \times 4 \times
4$ cube.  Define the sequence
\begin{equation}
\sigma = \{I, \sigma_z, \sigma_x, \sigma_x \sigma_z\}.
\end{equation}
Then, interpreting the index $k$ as an ordered tuple
$(k_0,k_1,\ldots,k_5)$, we have
\begin{equation}
C_{01k} = \hbox{Re}\,\tr(\Pi_0 \Pi_1 D_k \Pi_0 D_k^\dag),
\end{equation}
where
\begin{equation}
D_k = \sigma_{k_1 + 2k_0}
 \otimes \sigma_{k_3 + 2k_2}
 \otimes \sigma_{k_5 + 2k_4}.
\end{equation}
We can therefore display $C_{01k}$ for all $k$ in a three-dimensional
cube, which is portrayed in Figure~\ref{fig:triple-product-cube}.

The pairing of values follows from the facts that $C_{01k} = C_{10k}$
by symmetry and
\begin{equation}
D_1^2 = (I \otimes I \otimes Z)^2 = I \otimes I \otimes I.
\end{equation}
This makes the triple product insensitive to a $Z$
factor on one qubit.  However, if the displacement operator includes a
factor of~$X$ on that qubit, then the triple product $C_{01k}$
vanishes.  Inspection reveals that among the nonvanishing values,
$C_{01k} = -1/27$ when the displacement operator $D_k$ includes only
factors of~$X$, apart from the third qubit, which is insensitive
to~$Z$.

Define the complex triple products
\begin{equation}
T_{jkl} = \braket{\psi_j}{\psi_k}
         \braket{\psi_k}{\psi_l}
         \braket{\psi_l}{\psi_j}
 = \tr(\Pi_j\Pi_k\Pi_l).
\end{equation}
Up until now, we have taken the real part of this quantity.  We can
instead scale by the magnitude to obtain a phase~\cite{LieAlgebra1}:
\begin{equation}
\tilde{T}_{jkl} = \frac{T_{jkl}}{|T_{jkl}|}
 = e^{i\theta_{jkl}}.
\end{equation}
It follows from the definition of $T_{jkl}$ that, in general,
\begin{equation}
e^{i\theta_{mjk}} e^{i\theta_{mkl}} e^{i\theta_{mlj}}
 = e^{i\theta_{jkl}}.
\end{equation}
For the Hoggar SIC, $\theta_{jkl}$ takes the values $0$, $\pi$ and
$\pm \pi / 2$.

Note that the definition of a SIC implies that
\begin{equation}
\braket{\psi_j}{\psi_k} = \frac{1}{\sqrt{d+1}} e^{i\theta_{jk}}
\end{equation}
for some angles $\theta_{jk}$.  This two-index object is related to
the three-index object $\theta_{jkl}$ by
\begin{equation}
e^{i\theta_{jkl}} = e^{i\theta_{jk}} e^{i\theta_{kl}}
e^{i\theta_{lj}}.
\end{equation}
With this relation, we can understand more about the triple products
$C_{jkl}$ using the following sneaky trick.  The operators $X$ and $Z$
are Hermitian, but $XZ$ is not.  We can fix this by defining
\begin{equation}
Y = iXZ,
\end{equation}
which is a Hermitian operator (and equal to the familiar Pauli matrix
$\sigma_y$).  A tensor-product operator like $X \otimes Z \otimes XZ$
will not be Hermitian, but $X \otimes Z \otimes Y$ will be.  So, by
introducing appropriate phase factors, we can fix up the
Weyl--Heisenberg displacement operators $D_k$ so that they are
Hermitian matrices.  The phase with which we modify $D_k$
includes a factor of~$i$ for every instance of~$Y$ in the tensor product:
\begin{equation}
\hat{D}_k = (-e^{i\pi/d})^{\#(Y)} D_k.
\end{equation}
These operators serve just as well for generating a SIC.

But notice: Our displacement operators are now Hermitian matrices,
that is, \emph{quantum observables,} and their expectation values are
real.  Conseqently, for any $\hat{D}_k$,
\begin{equation}
\bra{\psi_0} \hat{D}_k \ket{\psi_0} \in \mathbb{R}.
\end{equation}
In turn, this implies that
\begin{equation}
e^{i\theta_{0k}} = \pm 1.
\end{equation}

Denote by $S_0$ the set of all triples $(jkl)$ for which $C_{jkl}$
vanishes.  For these triples, it must be the case that $T_{jkl}$ is
pure imaginary.  Let us focus on the case $j = 0$, with $k \neq 0$ and
$l \neq 0$.  Here, the only place a factor of~$i$ can enter is the
middle:
\begin{equation}
e^{i\theta_{0kl}} = e^{i\theta_{0k}} e^{i\theta_{kl}}
e^{i\theta_{l0}}.
\end{equation}
The middle factor is the phase of the inner product
\begin{equation}
\braket{\psi_k}{\psi_l} = \bra{\psi_0} \hat{D}_k^\dag
 \hat{D}_l \ket{\psi_0}.
\end{equation}
This can yield an imaginary part for some values of~$k$ and $l$,
thanks to the phase factors we introduced to obtain Hermiticity.
Write $\{\cdot,\cdot\}$ for the symplectic form
\begin{equation}
\{a,b\} = a_1b_0 - b_1a_0.
\end{equation}
Then the phase we obtain is
\begin{equation}
(-i)^{\{(k_0,k_1),(l_0,l_1)\}
       + \{(k_2,k_3),(l_2,l_3)\}
       + \{(k_4,k_5),(l_4,l_5)\}}.
\label{eq:symplectic-phase}
\end{equation} 
If we fix the index $k$, say to
\begin{equation}
(k_0,k_1,k_2,k_3,k_4,k_5) = (0,0,0,0,0,1),
\end{equation}
then the phase contribution will be an imaginary number for exactly 32
of the 64 possible choices of the index $l$.  These are the values for
which $C_{01l} = 0$.

If we define the matrix
\begin{equation}
\Omega = I_{3\times3} \otimes
 \left(\begin{array}{cc}
       0 & -1 \\ 1 & 0
       \end{array}
 \right),
\end{equation}
then we can write the exponent in Eq.~(\ref{eq:symplectic-phase}) as
\begin{equation}
f(k,l) = k\, \Omega\, l^{\rm T},
\label{eq:symp-bilinear}
\end{equation}
where we are interpreting $k$ and $l$ as row vectors of six elements
each.  The matrix $\Omega$ is invertible and antisymmetric, so $f(k,l)$ is
a \emph{symplectic bilinear form.}

Let us consider again the three-index angle tensor $\theta_{jkl}$.  We
know that
\begin{equation}
e^{i\theta_{jkl}} = \pm i,\hbox{ for } (jkl) \in S_0.
\end{equation}
If $(mjk)$, $(mkl)$ and $(mlj)$ are three triples in~$S_0$, then
\begin{equation}
e^{i\theta_{jkl}} = \pm i.
\end{equation}
That is, $(jkl)$ must then be a member of~$S_0$, too.  On the other
hand, if $(mjk)$, $(mkl)$ and $(mlj)$ are all \emph{outside} of~$S_0$,
then $e^{i\theta_{jkl}}$ is the product of three real numbers, and so
it must be real itself.  Therefore, if  $(mjk)$, $(mkl)$ and $(mlj)$
are in the complement of~$S_0$, then so is $(jkl)$.

This means that $S_0$ qualifies as a \emph{two-graph.}  Much studied
in discrete mathematics, a two-graph can be defined~\cite{cameron1980}
as a set $T$ of triples such that
\begin{equation}
(pqr), (pqs), (prs) \in T \Rightarrow (qrs) \in T,
\end{equation}
and likewise for the complement of~$T$.

One application of two-graphs is generating sets of equiangular lines
in \emph{real} vector spaces.  Pick a point in~$\mathbb{R}^d$, and
draw a set of lines through it, such that any two meet at an angle
whose cosine is $\pm \alpha$ (with $\alpha \neq 0$).  For some triples
of those intersecting lines, the product of the cosines will be
negative, and for others, it will be positive.  The triples for which
the product is negative constitute a two-graph.  Going in the other
direction, any two-graph can be formulated in this way.

Notice what has happened here:  We started with a set of
\emph{complex} equiangular lines, the Hoggar SIC, and in considering
the additional symmetries that set enjoys above and beyond its
definition, we have arrived at \emph{real} equiangular lines.

This will happen again.

Two-graphs have been taxonomied to an extent, with the aid of the
classification theorem for finite simple groups.  Those two-graphs
with doubly transitive automorphism groups were classified by
Taylor~\cite{taylor1992}.  Our set $S_0$ is Taylor's example B.xi, the
two-graph on 64 vertices whose automorphism group contains $\stabgp$.

Knowing the automorphism group of this two-graph, we can find the
stabilizer of any pair of vertices.  This will be the subgroup whose
action leaves that pair fixed.  For example, automorphisms in the
stabilizer subgroup of the pair $(0,1)$ will send the triple $(01k)$
to the triple $(01k')$.  Taylor~\cite{taylor1992} observes that the
stabilizer of two points in a triple has orbits of length 6, 24 and 32
on the remaining points.  Combining this with Zhu's
observation~\cite{zhu-thesis} that two triples in the Hoggar SIC can
be mapped to each other by a symmetry operation if and only if they
have the same triple product, and we see a combinatorial origin of the
patterns we observed in Figure~\ref{fig:triple-product-cube}.

Given a two-graph $T$, one can construct a regular graph $G$ that
embodies its structure in the following manner~\cite{taylor1992}.
Copy over the list of vertices from~$T$ to~$G$.  Then, select a vertex
$v$ of the two-graph $T$, and draw the edges of~$G$ so that $u$ and
$w$ are neighbors whenever $(uvw) \in T$.  Let $A$ be the \emph{Seidel
  adjacency matrix} of the graph $G$.  This matrix is constructed so
that $A_{uw} = -1$ if $u$ and $w$ are adjacent, $A_{uw} = 1$ if they
are not, and $A_{uu} = 0$ on the diagonal.  Suppose that the smallest
eigenvalue of~$A$ is $\lambda$, and this eigenvalue occurs with
multiplicity $m$.  Then, $M = I - (1/\lambda) A$ is a symmetric,
positive definite matrix, and the rank of~$M$ will be the number $|A|$
of vertices in the graph minus the multiplicity $m$.  Consequently,
$M$ can be taken as the Gram matrix for a set of vectors
\begin{equation}
\{v_1,v_2,\ldots,v_{|A|} \},
\end{equation}
with each vector living in $\mathbb{R}^{|A|-m}$.

In our case, the matrix $A$ has only two eigenvalues: 7, with
multiplicity 36; and $-9$, with multiplicity 28.  This means that $M$
is the Gram matrix for a set of \emph{equiangular} lines (as it should
be, since we derived $G$ from a two-graph).

From the triple-product structure of the Hoggar SIC, we have arrived
at a set of 64 equiangular lines in~$\mathbb{R}^{36}$.

The numbers 28 and 36 will recur in the next developments.

\section{The Twin of the Hoggar SIC}
\label{sec:twin}
\begin{myverbbox}{\vbfourlines}
1110111011100001111011101110000111101110111000010001000100011110
1101110111010010110111011101001011011101110100100010001000101101
1011101110110100101110111011010010111011101101000100010001001011
0111011101111000011101110111100001110111011110001000100010000111
\end{myverbbox}
\begin{table*}[t]
\begin{center}
\vbfourlines
\end{center}
\caption{\label{tab:4lines} Four SIC representations of states from
  the twin Hoggar lines, written as bit sequences.}
\end{table*}

Now, we investigate the eight-dimensional analogue of what happens
when we minimize the Shannon entropy for qubit pure states.

The ``twin Hoggar SIC'' can be constructed by applying the
triple-Pauli displacement operators to the fiducial vector
\begin{equation}
\ket{\tilde{\psi}_0} \propto (-1-2i, 1, 1, 1,
                      1, 1, 1, 1)^{\rm T}.
\end{equation}
This is related to our original fiducial vector,
Eq.~(\ref{eq:hoggar-fiducial}), by complex conjugation.

In the SIC representation defined by the original Hoggar lines, the
vectors comprising the ``twin Hoggar SIC'' have $(8-1)8/2 = 28$
elements equal to zero, and the other $(8+1)8/2 = 36$ elements equal
to $1/36$~\cite{Szymusiak2015}.  Consequently, the Hoggar lines
provide a counterexample to the conjecture that the best upper bound
on the number of zero-valued entries in dimension $d$ is just $d$.
The bound $d(d-1)/2$ deduced from the Cauchy--Schwarz
inequality~\cite{appleby2011} is, actually, tight.  Furthermore, the
states of the twin Hoggar SIC minimize the Shannon entropy of their
SIC representations, as we discussed above.  One can, in fact, find
the the twin Hoggar SIC-set by testing all the states of the
form~(\ref{eq:Shannon-minimizer}) to see which ones satisfy the QBic
equation.

For any vector $p$ in the twin Hoggar SIC set,
\begin{equation}
\sum_j p(j)^3 = 36 \left(\frac{1}{36}\right)^2
 = \frac{1}{1296}.
\end{equation}
Eq.~(\ref{eq:qbic-hoggar-boxed}) then becomes
\begin{equation}
\frac{1}{1296} 
+ \frac{1}{3} \left[\sum_{S_+} p(j)p(k)p(l)
   - \sum_{S_-} p(j)p(k)p(l)
   \right]
 =
\frac{11}{648}.
\end{equation}
The bracketed sum must therefore equal
\begin{equation}
\left[\sum_{S_+} p(j)p(k)p(l)
   - \sum_{S_-} p(j)p(k)p(l)
   \right]
 = \frac{7}{144}.
\end{equation}

Furthermore, any product $p(j)p(k)p(l)$ that does not evaluate to zero
must equal
\begin{equation}
p(j)p(k)p(l) = \left(\frac{1}{36}\right)^3 = \frac{1}{46,656}.
\end{equation}
From this, we can calculate the net number of contributions that the
sums over $S_+$ and $S_-$ must make, if the state is to be valid:
\begin{equation}
\frac{\frac{7}{144}}{\frac{1}{46,656}}
 = 2,268 = 2^2 3^4 7 = \frac{3(|S_+| - |S_-|)}{2^4}.
\end{equation}

If $\pi_i$ and $\pi_j$ are two projectors in the twin set, then
\begin{equation}
\tr(\pi_i \pi_j) = d(d+1) \sum_k p_i(k) p_j(k) - 1
 = \frac{1}{d+1}.
\end{equation}
Therefore,
\begin{equation}
\sum_k p_i(k) p_j(k) = \frac{d+2}{d(d+1)^2}
 = \frac{5}{324}.
\end{equation}
Now, each element in $p_i$ is either 0 or $1/36$, and likewise
for~$p_j$.  Let $n$ denote the number of overlapping nonzero entries
in these two vectors.  We know that
\begin{equation}
n\left(\frac{1}{36}\right)^2 = \frac{5}{324},
\end{equation}
and so
\begin{equation}
n = 20.
\end{equation}
This result will be important for understanding the twin Hoggar SIC
using combinatorial design theory.

\section{Combinatorial Designs from the Twin Hoggar SIC}
\label{sec:twin-design}
We have a set of $d^2 = 64$ ``blocks,'' each one of which essentially
is a binary string of length 64.  And each block contains exactly 36
of the nonzero entries that a length-64 block could in principle
contain.  We can think of this as there being 64 ``points,'' and each
block contains 36 of them.  Table~\ref{tab:4lines} gives examples of
four such blocks.

If we fix $v = b = 64$ and $k = 36$, then
\begin{equation}
\lambda \cdot 63 = 36 \cdot 35\ \Rightarrow\ \lambda = 20.
\end{equation}
This is just what we found before when we calculated the number of
overlapping 1s in any pair of vectors in the twin Hoggar set.
Therefore, the twin Hoggar SIC defines a symmetric design.
Specifically, it is a ``2-(64,36,20) design.''

If we apply a \textsc{not} to each of our bit-strings, then we arrive
at a new design.  Generally, the \emph{complement} of a design is
found by replacing each block with its complement: The points that
were included in a block are now excluded, and vice versa.  The new
design has parameters
\begin{equation}
v' = v,\ b' = b,\ k' = v - k,\ r' = b - r,\ \lambda' = 
\lambda + b - 2r.
\end{equation}
The complement to our Hoggar design therefore satisfies
\begin{equation}
v' = b' = 64,\ k' = r' = 28,\ \lambda' = 12.
\end{equation}
Therefore, we can designate it a ``2-(64,28,12) design.''

The existence of a symmetric design with parameters
\begin{equation}
(v,k,\lambda) = (4u^2, 2u^2 - u, u^2 - u) 
\end{equation}
is known to be equivalent to the existence of a regular Hadamard
matrix possessing dimensions $4u \times 4u$.  Setting $u = 4$, we find
that the complement of the Hoggar design meets the Hadamard
criterion.  The incidence matrix of the design can be transformed into
a regular Hadamard matrix by simple substitutions.

The complement of the Hoggar design is equivalent to an
\emph{orthogonality graph} for the Hoggar SIC and its twin.  In an
orthogonality graph, vertices stand for states, and vertices are
linked by an edge if the corresponding states are orthogonal.  If a
point $V_i$ lies within block $B_j$, then the $i$\textsuperscript{th}
vector in the Hoggar SIC is orthogonal to the $j$\textsuperscript{th}
vector in the twin SIC.  This can be visualized as a bipartite graph
containing two sets of 64 vertices apiece, where each vertex in the
first set is linked to 28 vertices in the second set.

We can generate the Hoggar design in another way by the following
procedure.  Start with this Hadamard matrix:
\begin{equation}
H_2 = \left(\begin{array}{rrrr}
 -1 & 1 & 1 & 1 \\
 1 & -1 & 1 & 1 \\
 1 & 1 & -1 & 1 \\
 1 & 1 & 1 & -1
 \end{array}\right).
\end{equation}
Construct the tensor product of three copies of $H_2$:
\begin{equation}
H_6 = H_2 \otimes H_2 \otimes H_2.
\end{equation}
Then, use this to create an incidence matrix by replacing all the
entries that equal $-1$ with~0:
\begin{equation}
M = \frac{H_6 + 1}{2}.
\end{equation}
The resulting $64 \times 64$ array is the incidence matrix of the
Hoggar design, containing all the same rows as the (appropriately
renormalized) SIC representations of the twin set.  This ties us
firmly into the literature on combinatorial designs: The Hoggar design
is a \emph{symplectic design on 64 points.}\footnote{While these notes
  were in preparation, Szymusiak and S\l{}omczy\'nski updated an
  earlier arXiv paper of theirs with an independent derivation of this
  point~\cite{Szymusiak2015}.}  A symplectic design~\cite{kantor1975,
  parker1994, cameron2003}, denoted $\sympdes^\epsilon(2m)$ with $m$ a
positive integer and $\epsilon = \pm 1$, is a BIBD with
\begin{equation}
b = v = 2^{2m},\ k = 2^{2m-1} + \epsilon 2^{m-1},
\ \lambda = 2^{2m-2} + \epsilon 2^{m-1}.
\end{equation}
The object that we found by way of SIC-POVMs is exactly $\sympdes^1(2m)$
for~$m = 3$.  Symplectic designs for larger $m$ can be constructed by
taking the tensor product of $m$ copies of the Hadamard matrix $H_2$.

That is how to construct the symplectic designs $\sympdes^\pm(6)$, as
combinatorial geometries.  Does the matrix $H_2$ have a meaning in
quantum physics?  In fact, it does.  In qubit state space, a SIC is a
tetrahedron inscribed within the Bloch sphere.  Finding the
minimum-entropy pure states, as we did for the Hoggar SIC, they turn
out to form a second tetrahedron, dual to the first.  Together, the
two SICs constitute a stellated octahedron in the Bloch-sphere
representation.  Each projector in the new SIC is orthogonal to
exactly one of the four projectors in the original SIC.  Let
$J_{4\times4}$ be the $4 \times 4$ matrix whose entries are all 1.
Then, up to normalization, the SIC representations of the four new
projectors can be written as the rows of the matrix
\begin{equation}
\left(\begin{array}{cccc}
 0 & 1 & 1 & 1 \\
 1 & 0 & 1 & 1 \\
 1 & 1 & 0 & 1 \\
 1 & 1 & 1 & 0
 \end{array}\right) = J_{4\times4} - I_{4\times4}.
\end{equation}
This is clearly just the Hadamard matrix $H_2$, shifted and rescaled.
So, the structure of orthogonalities between the Hoggar SIC and its
twin is, essentially, the tensor product of three copies of the
analogous structure for a qubit SIC.

An automorphism of a symmetric design is a permutation of the points
that preserves the block structure, sending blocks to blocks.  The
symplectic designs admit \emph{2-transitive automorphism groups.}
That is, the automorphism group of a symplectic design
$\sympdes^\epsilon(2m)$ contains permutations that map any pair of
points to any other pair of points.  Furthermore, the automorphism
group of a design is 2-transitive for points if and only if it is
so for blocks as well.  Therefore, the automorphism group of a
symplectic design includes transformations that can map any pair of
blocks to any other pair of blocks.

The \emph{symmetric difference} of two sets is defined to be the set
of those elements contained in their union but not their
intersection.  For example,
\begin{align}
\{\hbox{John}, \hbox{Paul}, \hbox{George}\}
\ominus \{\hbox{George}, \hbox{Ringo}\} \nonumber\\
\qquad\qquad = \{\hbox{John}, \hbox{Paul}, \hbox{Ringo}\}.
\end{align}
If the symmetric difference of any \emph{three} blocks in a design is
either a block or the complement of a block, then that design is said
to have the \emph{symmetric difference property.}  If a design enjoys
the symmetric difference property, then that design or its complement 
meets the following condition~\cite{dillon1987} on its parameters:
\begin{equation}
v = 2^{2m},\ k = 2^{2m-1} - 2^{m-1},
\ \lambda = 2^{2m-2} - 2^{m-1}.
\end{equation}
The complement of the Hoggar design satisfies these conditions with $m
= 3$.

The 64-point designs with the symmetric difference property can be
completely classified~\cite{dillon1987}.  There exist four
inequivalent such designs, distinguished by their automorphism
groups~\cite{parker1994}.  The symplectic design, which we found by
way of the Hoggar SIC, is the most symmetric: It is the only one of
the four whose automorphism group is 2-transitive.

Let $\mathbb{F}_2$ denote the finite field of order two, and
let $\hbox{Sp}(2m,\mathbb{F})$ denote the group of $2m\times 2m$
symplectic matrices over the field $\mathbb{F}$.  Then, the
automorphism group of the symplectic design $\sympdes^1(6)$ is
isomorphic to
\begin{equation}
G = (\mathbb{Z}_2)^6 \times \hbox{Sp}(6,\mathbb{F}_2).
\end{equation}
The stabilizer of any point is $\hbox{Sp}(6,\mathbb{F}_2)$.

The original SIC and the twin SIC have the same symmetry group.  Let
$\Pi_i$ be a projector in the original set and $\pi_i$ a projector in
the twin set.  Suppose that $g$ is an element of the symmetry group
that takes $\pi_j$ to $\pi_k$.  Then
\begin{equation}
\tr(\Pi_i \pi_k) = \tr(\Pi_i g \pi_j g^\dag) = \tr(g^\dag \Pi_i g \pi_j).
\end{equation}
So, the SIC representation of $\pi_j$ is just the SIC representation of
$\pi_k$, with the entries permuted.  Any element of the Hoggar SIC's
symmetry group corresponds to a permutation that preserves the
combinatorial design structure.  However, the converse is not true:
Not all elements in the automorphism group $G$ can be implemented by
unitaries that belong to the Hoggar SIC's symmetry group.  This is a
restatement of the fact that the symmetry group of the Hoggar SIC is a
proper subgroup of the triple-qubit Clifford group.

\section{Post-Peierls Compatibility}
\label{sec:PP}
A quantum state can be thought of as a hypothesis for how a quantum
system will behave when experimented upon.  When are two such
hypotheses different in a meaningful way?  One way of quantifying this
is the idea of \emph{compatibility} between quantum states.  Two
quantum states $\rho$ and $\rho'$ are \emph{post-Peierls incompatible}
if a measurement exists that meets the following
condition~\cite{CavesFuchsSchack2002}.  Let the measurement outcomes
be labeled by~$j$, so that the operators $\{E_j\}$ form a POVM,
\begin{equation}
\sum_j E_j = I.
\end{equation}
The probabilities for the outcomes are computed using the Born rule:
\begin{equation}
q(j) = \tr(\rho E_j),\ q'(j) = \tr(\rho' E_j).
\end{equation}
If one can devise a measurement $\{E_j\}$ such that for \emph{any}
outcome $j$, at least one of $q(j)$ or $q'(j)$ is zero, then the
states $\rho$ and $\rho'$ are post-Peierls (PP) incompatible.  This
can naturally be generalized to the question of compatibility among
three or more states.

Is it possible for quantum states to be PP incompatible with respect
to a SIC measurement?  Yes, but not if we only consider two states at
a time.  For example, these are three valid states for the Hesse SIC
representation in dimension $d = 3$.
\begin{equation}
\begin{array}{c}
\left(0,0,0,\frac{1}{6},\frac{1}{6},\frac{1}{6},
 \frac{1}{6},\frac{1}{6},\frac{1}{6}\right); \\
 \\
\left(\frac{1}{6},\frac{1}{6},\frac{1}{6},0,0,0,
 \frac{1}{6},\frac{1}{6},\frac{1}{6}\right); \\
\\
\left(\frac{1}{6},\frac{1}{6},\frac{1}{6},
 \frac{1}{6},\frac{1}{6},\frac{1}{6},0,0,0\right).
\end{array}
\label{eq:hesse-example}
\end{equation}
Note that there is exactly one zero in each column.  In other words,
for each outcome of the Hesse SIC, exactly one of these three states
assigns that outcome a probability of zero.

\begin{figure}[h]
\includegraphics[width=6cm]{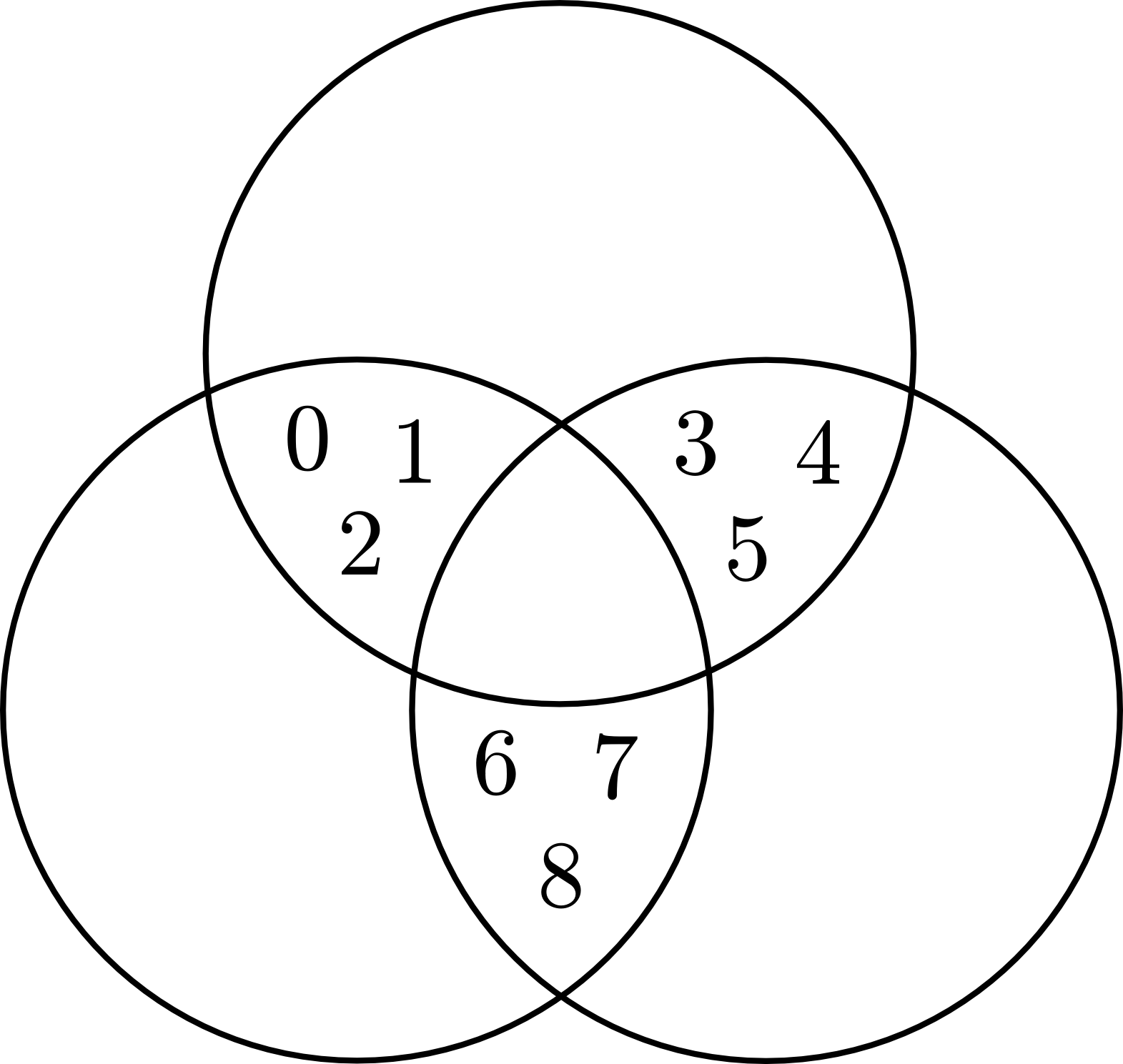}
\caption{\label{fig:venn-hesse} Pictorial representation of the
  hypotheses defined in Eq.~(\ref{eq:hesse-example}).  Each circle
  corresponds to a quantum state.  The numbers
  indicate the outcomes that are consistent with that state,
  \emph{i.e.,} the outcomes for which that state implies nonzero
  probability.}
\end{figure}

This is a situation where the relationship among three entities is not
clearly apparent from the relationships within each pair.  In such a
case, it can be helpful to portray the configuration
diagramatically~\cite{allen2014, stacey-thesis, complexityinf}.  We do
so in Figure~\ref{fig:venn-hesse}.  Each circle in
Figure~\ref{fig:venn-hesse} stands for one of the three states given
in Eq.~(\ref{eq:hesse-example}).  The numbers contained within a
circle are the labels of the outcomes that are consistent with that
state.  Note that these outcomes are only written in the areas where
two circles overlap.  No outcome belongs to a single state alone, and
no outcome belongs to all three.

Suppose we have three pure states in dimension $d = 3$.  We denote
them by $\ket{\psi}$, $\ket{\psi'}$ and $\ket{\psi''}$.  These can be
considered as three different hypotheses that an agent Alice is
willing to entertain about a quantum system.  If they are PP
incompatible, then there exists some measurement that Alice can
perform such that for any outcome of that measurement, at least one of
the three hypotheses deems that outcome impossible.  If we specialize
to \emph{von Neumann} measurements, then we can give a criterion for
``PP-ODOP'' compatibility (One-Dimensional, Orthogonal Projectors).  A
necessary and sufficient condition~\cite{CavesFuchsSchack2002,
  stacey-qutrit} for three pure states in~$d = 3$ to be PP-ODOP
incompatible is for the following inequalities to be satisfied.  First,
\begin{equation}
\left|\braket{\psi}{\psi'}\right|^2
 + \left|\braket{\psi'}{\psi''}\right|^2
 + \left|\braket{\psi''}{\psi}\right|^2 < 1, 
\label{eq:PP-ODOP-sym1}
\end{equation}
and second,
\begin{equation}
\begin{array}{c}
\left(\left|\braket{\psi}{\psi'}\right|^2
 + \left|\braket{\psi'}{\psi''}\right|^2
 + \left|\braket{\psi''}{\psi}\right|^2 - 1\right)^2 \\
 \geq 4\left|\braket{\psi}{\psi'}\right|^2
     \left|\braket{\psi'}{\psi''}\right|^2
     \left|\braket{\psi''}{\psi}\right|^2.
\end{array}
\label{eq:PP-ODOP-sym2}
\end{equation}

Consider what happens if the three states are drawn from a SIC set.
No set of three vectors can span more than three dimensions, so even
though our states naturally live in a higher-dimensional Hilbert
space, we press forward and use the three-dimensional criterion.  In
that case,
\begin{equation}
\left|\braket{\psi}{\psi'}\right|^2
 = \left|\braket{\psi'}{\psi''}\right|^2
 = \left|\braket{\psi''}{\psi}\right|^2
 = \frac{1}{d+1}.
\end{equation}
The first inequality becomes
\begin{equation}
\frac{3}{d+1} < 1,
\end{equation}
and the second inequality becomes
\begin{equation}
\left(\frac{3}{d+1} - 1\right)^2
 \geq \frac{4}{(d+1)^3}.
\end{equation}
We can simplify the latter expression to
\begin{equation}
(d-2)^2 \geq \frac{4}{d+1}.
\end{equation}
Both inequalities are satisfied simultaneously for~$d \geq 3$.

The three-dimensional criterion tells us that there is \emph{some} von
Neumann measurement with respect to which the three states drawn from
the SIC set are incompatible.  Yet the SIC is itself a measurement,
and with respect to \emph{that} measurement, the three states are
entirely compatible.  In the representation that the SIC itself
defines, the states $\ket{\psi}$, $\ket{\psi'}$ and $\ket{\psi''}$ all
have the form
\begin{equation}
e_k(i) = \frac{1}{d(d+1)} + \frac{1}{d+1} \delta_{ik}
\end{equation}
for some values of~$k$, and these vectors contain no zeros at all.
We have here a rather cute situation.  Classically, an
``informationally complete measurement'' would be something like an
experiment that discovers the exact values of all the positions and
momenta of the particles comprising a system.  We tend to think of any
other measurement as a coarse-graining of that one, a measurement that
throws away some of the information that is, in principle, available.
And throwing away information makes classical configurations
\emph{harder} to distinguish from one another.  If we can rule out a
hypothesis using a clumsy, imprecise measurement, then \emph{surely}
we could do so using a \emph{maximally informative} one!  How could a
measurement that is less exhaustive be \emph{better} at ruling out a
hypothesis?

This is indicative of the way in which quantum physics runs counter to
classical intuition.  An informationally complete quantum measurement
is \emph{not} the determination of the values of all hidden variables,
or the narrowing of a Liouville density to a delta function.  A vector
in a SIC representation is not a probability distribution over a
putative hidden-variable configuration space.  And we do not calculate
the probabilities for outcomes of other experiments merely by blurring
over IC ones.

The double-transitivity of the Hoggar SIC simplifies the structure of
the triple products, as we saw above.  It does the same for 
considerations of PP compatibility~\cite{CavesFuchsSchack2002}, as
well.

Any two SIC vectors are PP compatible.  However, a set of three SIC
vectors when taken together can be PP incompatible.  In dimension 3,
the measurements that reveal PP incompatibility for the Hesse SIC are
a collection of vectors originally known for other reasons: They
comprise four Mutually Unbiased Bases (MUB)~\cite{stacey-qutrit}.
What about with the Hoggar SIC?

Use one set of Hoggar lines to define a SIC representation of state
space, and translate the twin Hoggar lines into this
representation. Any two projectors in the twin Hoggar set will be
pairwise PP compatible.  Direct computation shows that any set of
three distinct projectors will also be compatible, in the sense that
the Hoggar SIC measurement itself will not reveal any incompatibility.
However, a set of four lines from the twin Hoggar set can be PP-POVM
incompatible, with that incompatibility revealed by the original
Hoggar SIC-POVM itself.  We will refer to this as ``PP-H
incompatibility.''  For example, in Table~\ref{tab:4lines} we gave the
SIC representations of four lines from the twin Hoggar set.

As we noted earlier, a ``1'' in a bitstring means that the entry in
that place is the nonzero value appropriate for the dimension, which
here is $1/36$.  A ``0'' means that the vector is zero in that slot.

There is at least one 0 in each column, meaning that for every
possible outcome of the Hoggar SIC-POVM, one of these four state
assignments deems that outcome impossible.  However, if we leave out
any of the four rows, this is no longer true.

\begin{figure}[h]
\includegraphics[width=4.5cm]{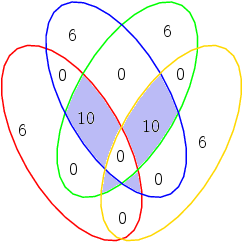}
\caption{\label{fig:hoggar-example} Venn diagram for the set of four
  states from the twin Hoggar SIC given in Table~\ref{tab:4lines}.
  Each ellipse stands for a quantum state.  Labels indicate the number
  of outcomes of the Hoggar SIC for which that state implies nonzero
  probability.  The shaded regions, where exactly three of the four
  ellipses overlap, contain 10 outcomes.  Each ellipse contains three
  such regions, as well as a region all to itself.  In total, each
  ellipse contains a value of 36.  The central region, where all four
  ellipses intersect, contains 0. (Figure based on~\cite{venn}.)}
\end{figure}

We illustrate this in Figure~\ref{fig:hoggar-example}.  Each ellipse
stands for a state vector, that is, for a row in
Table~\ref{tab:4lines}.  The central region, where all four ellipses
overlap, contains no outcomes.

There are lots of other examples; we do, after all, have 64 vectors to
choose from.  However, we should be able to simply the problem, and
understand what's going on by considering only a subset of all those
combinations.  Why?  

If we apply the same permutation to the four rows shown above, the
columns still line up, meaning that there is still at least one zero
in each column.  Consequently, the transformed states will also be
PP-H incompatible.

Because we can take any distinct pair $(\pi_j, \pi_k)$ to the pair
$(\pi_0, \pi_1)$, then we should be able to understand the PP-H
compatibility properties of all quadruples by working out what happens
with $(\pi_0, \pi_1, \pi_m, \pi_n)$.

We now apply our knowledge of combinatorial design theory.  Let $B_i$
denote the bitstring representation of the state $\pi_i$ in the twin
Hoggar SIC.  These 64 sequences, which we can think of as the rows in
a square matrix, form a symmetric design, as we showed earlier, and
this design has the symmetric difference property.  In terms of
bitstrings, the symmetric difference $B_i \ominus B_j$ is equivalent
to an \textsc{xor} operation:
\begin{equation}
(B_i \ominus B_j)(n) = B_i(n)\ \textsc{xor}\ B_j(n).
\end{equation}
This is readily verified, and implies the convenient fact that the
symmetric difference is associative:
\begin{equation}
(B_i \ominus B_j) \ominus B_k
 = B_i \ominus (B_j \ominus B_k).
\end{equation}
In Table~\ref{tab:xor}, we show the values resulting from applying
\textsc{xor} symmetrically to three bits, and the complementary values.

\begin{table}[h]
\begin{tabular}{ccc|c|c}
$a$ & $b$ & $c$ & $a\ \textsc{xor}\ b\ \textsc{xor}\ c$
 & $\textsc{not}(a\ \textsc{xor}\ b\ \textsc{xor}\ c)$ \\
\hline
0 & 0 & 0 & 0 & 1 \\
0 & 0 & 1 & 1 & 0 \\
0 & 1 & 0 & 1 & 0 \\
0 & 1 & 1 & 0 & 1 \\
1 & 0 & 0 & 1 & 0 \\
1 & 0 & 1 & 0 & 1 \\
1 & 1 & 0 & 0 & 1 \\
1 & 1 & 1 & 1 & 0
\end{tabular}
\caption{\label{tab:xor} The \textsc{xor} of three bits, and its
  complement.}
\end{table}

Because the Hoggar design has the symmetric difference property, the
symmetric difference of any three blocks is either a block or the
complement of a block.  Suppose that the symmetric difference of
$B_i$, $B_j$ and $B_k$ is the complement of~$B_l$.  Then $B_l$ is the
complement of $B_i \ominus B_j \ominus B_k$.  We can find each element
of~$B_l$ by locating the proper row in Table~\ref{tab:xor}.  It
follows that for all $n \in \{0,\ldots,63\}$, the set
\begin{equation}
\{ B_i(n), B_j(n), B_k(n), B_l(n) \}
\end{equation}
contains either 1, 2 or 3 zeroes.  That is, these elements are never
all zero, nor are they all ever one.  Consequently, a measurement of
the Hoggar SIC-POVM reveals PP-H incompatibility among the four states
$\{\pi_i, \pi_j, \pi_k, \pi_l\}$ in the twin Hoggar SIC.

What else can we say about the symmetric differences of the blocks
$\{B_i\}$?  Each $B_i \ominus B_j$ for a distinct pair $i \neq j$ is a
list of positions where exactly one of $B_i(n)$ and $B_j(n)$ equals
one.  By direct computation, we find that each such list is 32 items
long.  We can pick a pair of distinct blocks in 2,016 different ways.
However, not all choices yield different lists of positions.  In fact,
only 126 lists occur.  This is a consequence of a result noticed by
Kantor~\cite{kantor1975}: The symmetric differences in the symplectic
designs $\sympdes(2m)$ correspond to the \emph{hyperplanes} in the
$2m$-dimensional discrete affine space on the finite field of order 2,
denoted $\mathbb{F}_2$.  In the case $m = 3$, there are 126 such
hyperplanes, each containing $2^5 = 32$ points.  Each hyperplane is
the symmetric difference of 16 different choices of block pairs.

From Kantor's work, we can also extract a criterion for when a set of
three blocks $\{B_i, B_j, B_k\}$ will be part of a PP-H-incompatible
quadruple.  As we deduced, this occurs when the symmetric difference
of the three blocks is the complement of a block.  The quantity
\begin{equation}
\left| (B_i \ominus B_j) \cap B_k \right|
\end{equation}
equals either 16 or 20, depending on the choice of blocks.  When it
equals 16, the symmetric difference of the three blocks is itself a
block.  On the other hand, when it equals 20, then the symmetric
difference is the complement of a block, and we have the
incompatibility we seek.  This can be interpreted in terms of another
affine space on the finite field $\mathbb{F}_2$.  In this space, the
points are the 64 bitstrings of the twin Hoggar SIC.  For a fixed
$B_i$ and $B_j$ with $j \neq i$, the set of all $B_k$ such that 
\begin{equation}
\left| (B_i \ominus B_j) \cap B_k \right| = 16
\end{equation}
defines a hyperplane in this affine space.  Points that lie outside
this hyperplane correspond to bitstrings which, together with $B_i$
and $B_j$, can form part of a PP-H-incompatible quartet.

This construction also tells us about the triple products, in a way
that relates back to our symplectic bilinear form,
Eq.~(\ref{eq:symp-bilinear}).  Consider the quartet formed by~$B_i$,
$B_j$, $B_k$ and their symmetric difference.  If this quartet is PP-H
incompatible, then
\begin{equation}
\hbox{Re}\,\tr(\Pi_i \Pi_j \Pi_k)
 = \hbox{Re}\,\tr(\pi_i \pi_j \pi_k)
 = 0.
\end{equation}
In dimension 3, the triple products of the Hesse SIC depend on whether
or not three points are collinear~\cite{stacey-qutrit}.  Now, we see
that in dimension 8, triple products depend upon whether three points
lie in the same hyperplane.

In fact, the implication works both ways: If $(ijk) \in S_0$, then
$B_i$, $B_j$ and $B_k$ can be extended to form a PP-H-incompatible quartet.

\section{Deeper Into the Bitstrings}
\label{sec:deeper}
The bit $B_j(n)$ will be 0 if the inner product
\begin{equation}
\tr(\Pi_n \pi_j) = \tr(D_n \Pi_0 D_n^\dag D_j \pi_0 D_j^\dag)
\end{equation}
vanishes.  Here, the displacement operators $D_n$ and $D_j$ are built
from tensor products of the Pauli matrices.  Note that we can use the
cyclic property of the trace to reduce the problem to investigating
inner products of the form
\begin{equation}
\tr(\Pi_0 D_m \pi_0 D_m^\dag).
\end{equation}
The product $\Pi_0 \pi_0$ is a symmetric matrix.  If we want the trace
to vanish, we should try introducing an asymmetry somehow.

Of the four Pauli matrices, three (counting the identity) are
symmetric.  Only $Y$, which is proportional to the product
$X Z$, is antisymmetric.  We therefore make the educated
guess that the inner product will vanish if the displacement operator
$D_m$ involves \emph{an odd number of factors} of the Pauli matrix
$Y$.  This happens in 28 out of the 64 possible displacement
operators $D_m$, which is the number we're looking for.  Why 28?  If
we want one factor of~$Y$, we have three places to put it, and
we have $3^2 = 9$ choices for the other two factors.  This gives us 27
possible operators.  Then, the operator $YYY$ is also antisymmetric,
making a total of 28.

It is straightforward to check that these zeros fall in the correct
places to reproduce the first row of Table~\ref{tab:4lines}.

The displacement operator $D_m$ will be antisymmetric if a certain sum
has odd parity:
\begin{equation}
m_0 m_1 + m_2 m_3 + m_4 m_5 = 1 \mod 2.
\label{eq:odd-parity}
\end{equation}
This construction for picking 28 configurations out of 64 also arises
in the study of \emph{bitangents} to \emph{quartic
  curves}~\cite{gray1982}.  Take the plane $\mathbb{R}^2$, and define
a curve on the plane by a fourth-degree equation in two variables.
Such a curve can have as many as 28 bitangent lines, \emph{i.e.,}
lines that are tangent to the curve at exactly two places.  By
extending to the complex projective plane, one can always find a full
set of 28 bitangents.  Each one is labeled by a set of binary
coordinates satisfying Eq.~(\ref{eq:odd-parity}).

Rather unexpectedly, then, the study of SICs has made contact with the
theory of algebraic curves!

Consider the elements of the index $k$ that indicate the powers to
which we raise $X$ when constructing $D_k$, that is, the ordered triple
$(k_0, k_2, k_4)$.  This triple can take eight different values, seven
of them nonzero.  Likewise, we have seven nonzero possibilities for
$(k_1, k_3, k_5)$.  Let us group the possibilities for these two
ordered triples according to when the dot product has even parity:
\begin{equation}
k_0 k_1 + k_2 k_3 + k_4 k_5 = 0 \mod 2.
\label{eq:even-parity}
\end{equation}
For each choice of $(k_1, k_3, k_5)$, there are three choices for
$(k_0, k_2, k_4)$ that satisfy Eq.~(\ref{eq:even-parity}).
\begin{equation}
\begin{array}{cc}
(k_0, k_2, k_4) & (k_1, k_3, k_5) \\
010, 011, 001 & 100 \\
001, 101, 100 & 010 \\
010, 110, 100 & 001 \\
001, 111, 110 & 110 \\
010, 111, 101 & 101 \\
011, 111, 100 & 011 \\
110, 101, 011 & 111
\end{array}
\end{equation}

This configuration has a name.  The choices for $(k_0, k_2, k_4)$
label the points of the \emph{Fano plane,} and $(k_1,k_3,k_5)$ label
the lines.  The Fano plane has seven points and seven lines.  Each
point lies on three lines, and each line contains three points.  A
line and a point of the Fano plane are incident if and only if their
coordinates satisfy Eq.~(\ref{eq:even-parity}).

In the Fano plane, there are 28 ways to select a point and a line
\emph{not} incident with it: For each point, four of the seven lines
do not go through that point, and we have seven ways to choose a
point.  In discrete geometry, a \emph{flag} is the combination of a
line and a point lying on that line, and an \emph{anti-flag} is a line
with a point lying off that line.  So, there are 28 anti-flags in the
Fano plane, and for each of them, the dot product of the point and
line labels has \emph{odd} parity.  That is, for each anti-flag, the
label of the point and the label of the line satisfy
Eq.~(\ref{eq:odd-parity}).

Look back at Table~\ref{tab:4lines}.  Each occurrence of the bit 0 is
an anti-flag in a Fano plane!  We use the powers to which we raise $X$
to pick a point, and the powers to which we raise $Z$ to pick a line
(or vice versa).  If the point lies off the line, we write a 0.  All
other bits in the sequence, we set to~1.

We have not yet exhausted the numerology of the integer 28.  The
bitangents to a quartic curve can also be
identified~\cite{coxeter1988, green2013} with \emph{pairs of opposing
  vertices} in the \emph{Gosset polytope} $3_{21}$, an object living
in~$\mathbb{R}^7$ that is related to the Lie algebra $E_7$.  We can
construct this polytope in the following way~\cite{baez2013}.  Start
with the two vectors
\begin{equation}
(3,3,-1,-1,-1,-1,-1,-1) \hbox{ and }
(-3,-3,1,1,1,1,1,1),
\end{equation}
which both live in~$\mathbb{R}^8$.  Permute the entries of these
vectors in all possible ways.  This creates 56 vectors
in~$\mathbb{R}^8$.  All of them are orthogonal to the vector
\begin{equation}
(1,1,1,1,1,1,1,1),
\end{equation}
so they actually all fit into~$\mathbb{R}^7$.  These are the vertices
of the Gosset polytope.  Each pair of opposite vectors defines a line
through the origin, yielding 28 lines\ldots\ which turn out to be
equiangular.

We have here another unforeseen relation between the complex and the
real versions of the equiangular lines question.  Starting with one
maximal set of complex equiangular lines, we construct another.  The
fact of a vector in one set being orthogonal to a vector in the other
corresponds to a \emph{real} line in a maximal equiangular set thereof.

\section{Concluding Remarks}

SICs are a confluence of multiple topics in mathematics.
Weyl--Heisenberg SIC solutions in dimensions larger than 3 turn out to
have deep number-theoretic properties, connecting quantum information
theory to Hilbert's twelfth problem~\cite{RCF-SIC}.  The other known
SIC solutions, which we have termed the sporadic SICs, relate by way
of group theory to sphere packing and the
octonions~\cite{stacey-sporadic}.  By asking a physicist's
question---``Given this constraint, which states maximize and minimize
the entropy?''---we launched ourselves into symplectic designs,
two-graphs, bitangents to quartic curves and Gosset polytopes.
Prolonged exposure to the SIC problem makes one suspect that the
interface between physics and mathematics does not have the shape that
one first expected.

For each of the SICs with doubly transitive symmetry groups, the pure
states that minimize the Shannon entropy of the SIC representation are
related to equiangular real lines. In dimension 2, they form a
SIC~\cite{Szymusiak2015}, which is a tetrahedron in the Bloch ball,
and that yields four real lines. In dimension 3, they form 12 MUB
states~\cite{stacey-qutrit}. Picking one state from each MUB, we
obtain four equiangular lines in nine-dimensional real space. (There
are 81 ways to do this.)  And in dimension 8, the procedure yields the
twin Hoggar SIC, which is equivalent to 64 equiangular lines
in~$\mathbb{C}^8$. Furthermore, when we consider the relation between
the original SIC and its twin, we find a set of 28 lines, which are a
maximal set for 7- or 8-dimensional real vector space.  And, as we
remarked before, the triple-product structure of the Hoggar SIC leads
to a two-graph on 64 vertices, which is itself equivalent to a set of
equiangular lines in~$\mathbb{R}^{36}$.

That the solutions to the real and complex versions of the equiangular
lines problem should be related in this way is rather surprising.

To draw this essay to a close, we should note that the Hoggar SIC
provides a rather clean and elementary introduction to several
mathematical structures that have been employed in the study of
three-qubit quantum systems~\cite{cerchiai2010}.  For example, we
encountered the group $\stabgp$: It was (up to isomorphism) simply the
group of transformations that permute the vectors in the Hoggar SIC
while leaving the fiducial untouched.  This group has also
appeared~\cite{planat2013} in studies of Bell--Kochen--Specker
phenomena, that is, of the nonclassical meshing together of
probability assignments~\cite{RMP, Fuchs2014, stacey-qutrit,
  mermin1993, mermin1993-erratum}.  Likewise, the sorting of tensor
products of Pauli operators into symmetric and antisymmetric matrices
has been invoked in other problems~\cite{levay2008}.  We remarked upon
the appearance of a polytope related to an exceptional Lie algebra;
this, too, is a type of structure pertinent to Bell--Kochen--Specker
phenomena in three-qubit systems~\cite{waegell2015, loveridge2015}.
All this suggests that more ideas might yet be grown from the Hoggar
SIC.


\end{document}